\documentclass[showpacs,amsmath,superscriptaddress,reprint,aps]{revtex4-1}

\usepackage{graphicx}
\usepackage{hyperref}

\begin{document}

\title
{Enhanced single photon emission from carbon nanotube dopant states coupled to silicon microcavities}

\author{A.~Ishii}
\affiliation{Nanoscale Quantum Photonics Laboratory, RIKEN, Saitama 351-0198, Japan}
\affiliation{Quantum Optoelectronics Research Team, RIKEN center for Advanced Photonics, Saitama 351-0198, Japan}
\author{X.~He}
\affiliation{Center for Integrated Nanotechnologies, Materials Physics and Applications Division, Los Alamos National Laboratory, Los Alamos, New Mexico 87545, United States}
\author{N.~F.~Hartmann}
\affiliation{Center for Integrated Nanotechnologies, Materials Physics and Applications Division, Los Alamos National Laboratory, Los Alamos, New Mexico 87545, United States}
\author{H.~Machiya}
\affiliation{Nanoscale Quantum Photonics Laboratory, RIKEN, Saitama 351-0198, Japan}
\affiliation{Department of Electrical Engineering, The University of Tokyo, Tokyo 113-8656, Japan}
\author{H.~Htoon}
\affiliation{Center for Integrated Nanotechnologies, Materials Physics and Applications Division, Los Alamos National Laboratory, Los Alamos, New Mexico 87545, United States}
\author{S.~K.~Doorn}
\affiliation{Center for Integrated Nanotechnologies, Materials Physics and Applications Division, Los Alamos National Laboratory, Los Alamos, New Mexico 87545, United States}
\author{Y.~K.~Kato}
\affiliation{Nanoscale Quantum Photonics Laboratory, RIKEN, Saitama 351-0198, Japan}
\affiliation{Quantum Optoelectronics Research Team, RIKEN center for Advanced Photonics, Saitama 351-0198, Japan}
\email[Corresponding author. ]{yuichiro.kato@riken.jp}

\begin{abstract}
Single-walled carbon nanotubes are a promising material as quantum light sources at room temperature and as nanoscale light sources for integrated photonic circuits on silicon. Here we show that integration of dopant states in carbon nanotubes and silicon microcavities can provide bright and high-purity single photon emitters on silicon photonics platform at room temperature. We perform photoluminescence spectroscopy and observe enhancement of emission from the dopant states by a factor of $\sim$100, and cavity-enhanced radiative decay is confirmed using time-resolved measurements, where $\sim$30\% decrease of emission lifetime is observed. Statistics of photons emitted from the cavity-coupled dopant states are investigated by photon correlation measurements, and high-purity single photon generation is observed. Excitation power dependence of photon emission statistics shows that the degree of photon antibunching can be kept low even when the excitation power increases, while single photon emission rate can be increased up to $\sim 1.7 \times 10^7$~Hz.
\end{abstract}

\maketitle

Single photon emitters are a fundamental element for quantum information technologies \cite{OBrien:2009}, and a wide range of materials has been explored to obtain ideal single photon emitting devices \cite{Aharonovich:2016}. In particular, semiconducting carbon nanotubes (CNTs) are regarded as a promising material for such an application because they are a nanoscale light emitting material \cite{Avouris:2008} having stable excitonic states which arise from the one-dimensional structure of CNTs \cite{Wang:2005, Maultzsch:2005}. Under cryogenic temperatures, excitons in CNTs are localized and behave as quantum-dot-like states \cite{Hofmann:2016}, exhibiting a quantum light signature \cite{Hogele:2008, Hofmann:2013}. At room temperature, single photon generation using CNTs has already been accomplished by two approaches \cite{Endo:2015, Ma:2015NatNano, He:2017, Ma:2015PRL, Ishii:2017}. The first is where exciton trapping sites are created to localize excitons \cite{Ma:2015NatNano, He:2017}, and the second is where efficient exciton-exciton annihilation process is used to reduce the number of mobile excitons to unity \cite{Ma:2015PRL, Ishii:2017}. The approach using exciton trapping sites allows for high purity single photon generation, use of chirality sorted CNTs, and direct deposition on various types of substrates. Furthermore, trapping sites protect excitons from quenching sites in CNTs, and optically-allowed defect states appear below the dark states of $E_{11}$ excitons, resulting in significant brightening of photoluminescence \cite{Piao:2013, Miyauchi:2013}. Recently, aryl $sp^3$ defects have received considerable attention because of the wide range of selectability of CNT chiralities, dopant species, and reaction conditions, which allows for tunable emission wavelength and decay lifetime \cite{Hartmann:2016}. Using this method, single photon generation with a purity of 99\% and an emission wavelength of 1550~nm is achieved at room temperature \cite{He:2017}.

For practical single photon sources, not only single photon purity and operating temperature, but also emission wavelength, linewidth, brightness, and photon extraction efficiency are important. From this aspect, cavity structures are widely used to improve performances of single photon emitters \cite{Strauf:2007, Englund:2010, Kaupp:2016}. As for CNT single photon emitters, photonic \cite{Newman:2012, Jeantet:2016} and plasmonic \cite{Luo:2017} cavity configurations have been used to enhance the brightness of single photon emission at low temperature. Further development is expected by integrating single photon emitters into silicon photonics because it can lead to on-chip integrated quantum devices \cite{Harris:2016}, and CNTs have a potential for such an application due to their emission wavelengths having low transmission losses in silicon. Microcavities on silicon substrates have been used to enhance photoluminescence (PL) \cite{Watahiki:2012, Noury:2015} and Raman \cite{Sumikura:2013} signal, and efficient coupling even to a single carbon nanotube has also been achieved \cite{Imamura:2013, Miura:2014, Machiya:2018}, demonstrating that CNTs are suitable for integration with silicon photonics.

Here we report on integration of CNT dopant state emitters with silicon microcavities. Emission from aryl $sp^3$ defect states in CNTs coupled to two-dimensional photonic crystal microcavities is characterized by PL microscopy, and significant enhancement of PL intensity is observed. Time-resolved PL measurements on the same device show a direct evidence of enhanced emission decay rates by the Purcell effect, and we confirm single photon emission from the device by performing photon correlation measurements. The zero-delay second-order autocorrelation $g^{(2)}(0)$ is as low as 0.1, showing high-purity single photon generation, and the value is stable even at high power excitation, which allows for single photon emission rates as high as $\sim 1.7 \times 10^7$~Hz.

\begin{figure}
\includegraphics{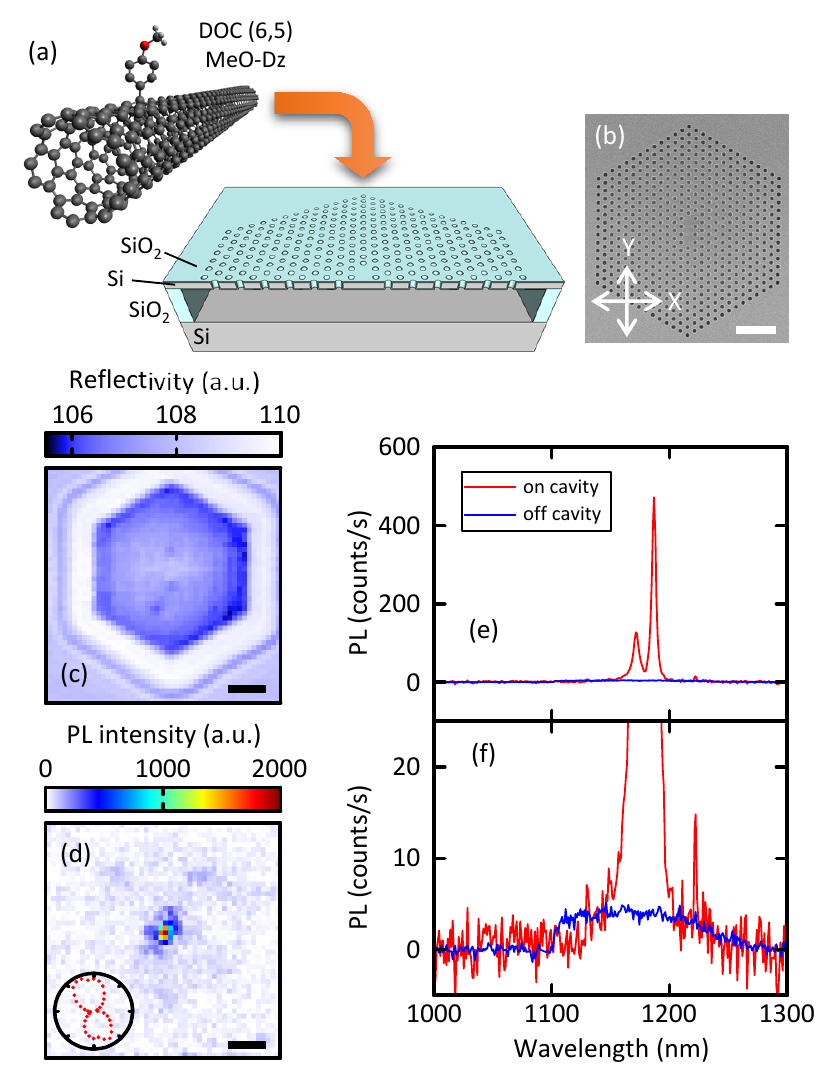}
\caption{
\label{Fig1} (a) Schematic images of a doped CNT and a photonic crystal microcavity. (b) A scanning electron micrograph of a photonic crystal microcavity. The arrows define the directions of $X$- and $Y$-polarization for excitation. (c) and (d) Reflectivity and PL images, respectively. Inset in (d) is the laser polarization dependence of PL intensity on the cavity. For (d) and the inset in (d), PL intensity is obtained by integrating PL over a 10-nm wide spectral window centered at 1187~nm. (e) PL spectra taken on the cavity (red) and off the cavity (blue). (f) An enlarged view of the low-intensity region of the data shown in (e). (b-d) The scale bars are 2~$\mu$m. (c-f) $Y$-polarized CW laser with $P=1$~$\mu$W is used for excitation. (e) and (f) The long-pass filter with a cut-on wavelength of 1100~nm is used when the off-cavity spectrum is taken.}
\end{figure}

We start sample preparation by fabrication of photonic crystal microcavities on a silicon-on-insulator substrate [Fig.~\ref{Fig1}(a)]. Electron beam lithography defines the photonic crystal pattern with shift-L3 cavities \cite{Akahane:2003Nature}, and the 200-nm-thick top Si layer is etched through by dry etching. The buried SiO$_2$ layer with a thickness of 1000~nm is then etched by 20~wt\% hydrofluoric acid, and thermal oxidation is performed at 900$^\circ$C for an hour to form a 10-nm-thick SiO$_2$ layer on the top Si layer. A scanning electron micrograph of a typical device is shown in Fig.~\ref{Fig1}(b). Doped carbon nanotubes are prepared from chirality-enriched (6,5) CNTs encapsulated in a sodium deoxycholate (DOC) surfactant. Aryl functionalization is done using a diazonium dopant (4-methoxybenzenediazonium (MeO-Dz)), where the details are described in the literature \cite{Hartmann:2015}. We dilute the doped CNT solution with water to avoid bundling or piling up of CNTs on a substrate, and finally the solution is drop casted on the devices using a glass micropipette.

PL measurements are performed with a homebuilt sample-scanning confocal microscopy system \cite{Ishii:2017}. We use a Ti:sapphire laser where the output can be switched between continuous-wave (CW) and $\sim$100-fs pulses with a repetition rate of 76~MHz. We use an excitation wavelength of 855~nm, which matches the phonon side-band absorption for (6,5) CNTs \cite{Vora:2010}. The excitation laser beam with a power $P$ is focused onto the sample by an objective lens with a numerical aperture of 0.85. PL and the reflected beam are collected by the same objective lens and separated by a dichroic filter. A Si photodiode detects the reflected beam for imaging, while a translating mirror is used to switch between PL spectroscopy and time-resolved PL measurements. PL spectra are measured with an InGaAs photodiode array attached to a spectrometer. For time-resolved measurements, $E_{11}$ emission at around 1000~nm is filtered out by a wavelength-tunable band-pass filter with a transmission window of 10~nm or a long-pass filter with a cut-on wavelength of 1100~nm. Fiber-coupled two-channel superconducting single photon detector (SSPD) connected to a 50:50 signal-splitting fiber is used to perform PL decay and photon correlation measurements. All measurements are conducted at room temperature in a nitrogen-purged environment.

We perform automated collection of PL spectra \cite{Ishii:2015} at all cavity positions to find devices with good optical coupling. For a device where a significant enhancement of the dopant state (${E_{11}}^*$) emission \cite{Hartmann:2016} is observed, reflectivity and PL images are taken [Figs.~\ref{Fig1}(c) and (d)]. The enhanced PL is localized at the center of the cavity, as expected from PL enhancement due to coupling with the resonance modes of the cavity. In Figs.~\ref{Fig1}(e) and (f), PL spectra on and off the cavity are shown, where the off-cavity signal is taken on the photonic crystal pattern. PL spectrum on the cavity shows multiple modes coupled to the dopant state emitters, while a broad emission peak from the dopant state is observed at the off-cavity position. In the on-cavity spectrum, the peak showing the highest intensity at an emission wavelength of 1187~nm has a full-width at half-maximum of 3.9~nm, corresponding to a quality factor $Q = 300$. We assign the highest intensity peak to the 2nd mode of the L3 cavity \cite{Liu:2015}. By comparing the peak height of the on-cavity and off-cavity PL spectra, we obtain a PL enhancement factor of $\sim$100.

The PL enhancement can become large as there are other cavity-induced effects in addition to the Purcell effect. In our devices, it is known that localized guided modes can increase the excitation by more than a factor of 50 \cite{Liu:2015}, and coupling to such an absorption resonance can explain the strong excitation polarization dependence (inset of Fig.~\ref{Fig1}(d)). Furthermore, the directionality of the cavity radiation can improve the PL collection efficiency by as much as a factor of 4 \cite{Fujita:2005}. Combined with the Purcell effect, these cavity effects can significantly brighten the nanotube emitters, and thus the obtained enhancement factor of $\sim$100 would be a reasonable result.

\begin{figure}
\includegraphics{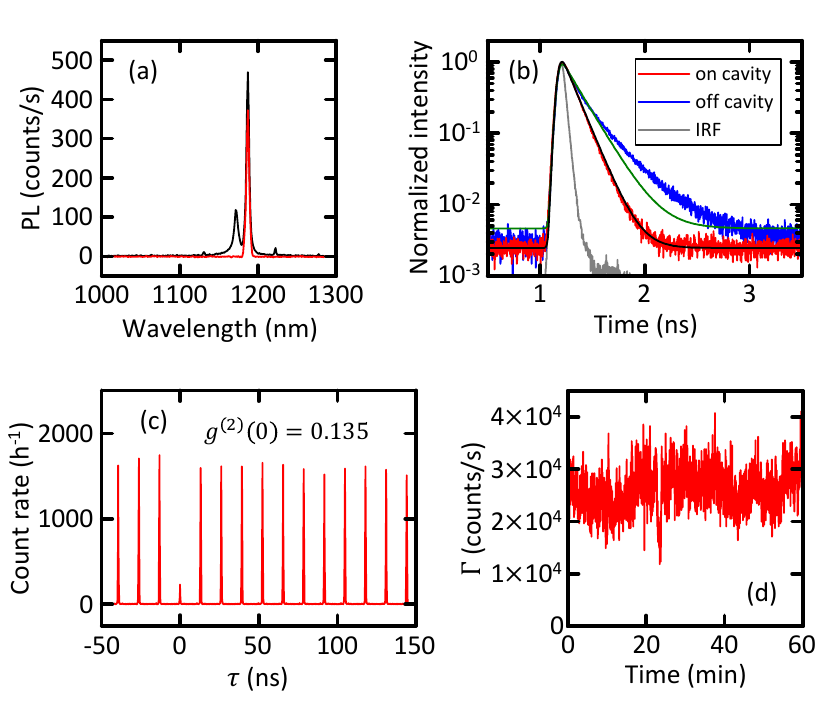}
\caption{
\label{Fig2} 
(a) PL spectra on the cavity taken before (black) and after (red) the band-pass filter is set. CW laser with $P=1$~$\mu$W is used for excitation. (b) PL decay curves taken with pulsed laser excitation at $P=0.1$~$\mu$W. The red and blue lines are for the on-cavity and typical off-cavity data, respectively. Fits with a convoluted mono-exponential decay function are also shown for the on the cavity (black curve) and off the cavity (green curve) data. Gray solid line represents the IRF. (c) and (d) An autocorrelation histogram and a time trace of photon detection rate, respectively, taken on the cavity at $P=0.5$~$\mu$W. An integration time of one second is used for each data point in (d). (a-d) $Y$-polarized laser is used for excitation. All the measurements are performed at room temperature.}
\end{figure}

In order to investigate the Purcell enhancement of the radiative decay rate, we perform time-resolved PL measurements on the same device shown in Figs.~\ref{Fig1}(c-f). For on-cavity PL, a single peak is spectrally filtered by tuning the transmission wavelength of the band-pass filter [Fig.~\ref{Fig2}(a)], while the long-pass filter is used instead for off-cavity PL. In Fig.~\ref{Fig2}(b), PL decay curves taken at the on-cavity and off-cavity positions are shown, and fits are performed using a mono-exponential decay function convoluted with a Gaussian profile representing the instrument response function (IRF) of the system. Although PL decay of doped CNTs typically exhibits a bi-exponential curve \cite{Hartmann:2016, He:2017}, here we use a mono-exponential decay for simplicity. From the fits, we obtain the on-cavity PL lifetime $\tau_\mathrm{on}=122.0\pm0.2$~ps and the off-cavity PL lifetime $\tau_\mathrm{off}=173.8\pm0.4$~ps. If we assume the radiative quantum efficiency $\eta$ of 2.4\%, which is estimated by the unaffected quantum efficiency of $\sim$11\% for MeO-Dz doped (6,5) CNTs in water \cite{Piao:2013} and PL quenching by a factor of $\sim$4.5 caused by an interaction with the SiO$_2$ substrate \cite{He:2017}, the $\sim$30\% reduction of the emission lifetime corresponds to a Purcell factor $F_p = (\tau_\mathrm{off}/\tau_\mathrm{on}-1) \eta^{-1}= 18$, a coupling factor $\beta=F_p/(1+F_p) = 0.95$, and an enhanced radiative quantum efficiency of 31\%.

We perform such lifetime measurements on ten other devices and obtain an average lifetime of 131.2$\pm$43.8~ps. The lifetime varies on different devices, suggesting that coupling is affected by some uncontrolled factors. Coupling efficiency is in general affected by spectral overlap, spatial overlap, and polarization overlap between the emitters and cavity modes. Moreover, CNT density fluctuations may have a significant effect in our samples. We note that variations in the cavity quality factors are not the main reason because $Q$ of dopant state emission is much lower than that of the cavity peak.

Next we measure photon correlation for the same cavity peak shown in Figs.~\ref{Fig2}(a) and (b), and clear photon antibunching is observed as shown in Fig.~\ref{Fig2}(c). We evaluate the normalized second-order autocorrelation at zero time delay $g^{(2)}(0)$ from the autocorrelation histogram by subtracting the dark counts and binning each peak with a binning width of 2~ns. We note that the histogram is taken within a time window from $-$60 to 300~ns, and side peaks after $\tau = 60$~ns are used for normalization to avoid underestimation due to photon bunching. For the data shown in Fig.~\ref{Fig2}(c), we obtain $g^{(2)}(0) = 0.136\pm0.005$, indicating high-purity single photon emission.

It is surprising that we obtain such high-purity single photon emission from a sample with drop-casted CNTs, where numerous emitters are expected within the laser spot. One explanation is that cavity coupling and spectral filtering allow selective photon collection from a few number of emitters. In fact, we actually observe higher $g^{(2)}(0)$ when the band-pass filter is removed [Supporting Information S1]. It is worth mentioning that we could not measure photon correlation of off-cavity signal as the emission intensity is too low, indicating the advantages of cavity coupling.

During the photon correlation measurements, time traces of the photon detection rate are also recorded [Fig.~\ref{Fig2}(d)]. The total photon detection rate $\Gamma$, defined as the sum of detection rates at the two channels, is obtained from the autocorrelation count rate $C$ using the relation
\begin{equation}
\Gamma = \left( \sqrt{r} + \frac{1}{\sqrt{r}} \right) \sqrt{\frac{C}{T}},
\end{equation}
where $r$ is the signal splitting ratio between the two channels, and $T=353.7$ ns is the effective time window for the 27 peaks which are included in the correlation histograms [Supporting Information S2]. In Fig.~\ref{Fig2}(d), the PL intensity shows a relatively large fluctuation over time, whose standard deviation is $\sim$23 times larger than that of shot-noise limited fluctuation. We observe such intensity fluctuation of the cavity-coupled peak for all devices we have measured, which may be caused by the influence of the substrate \cite{Newman:2012}.

\begin{figure*}
\includegraphics{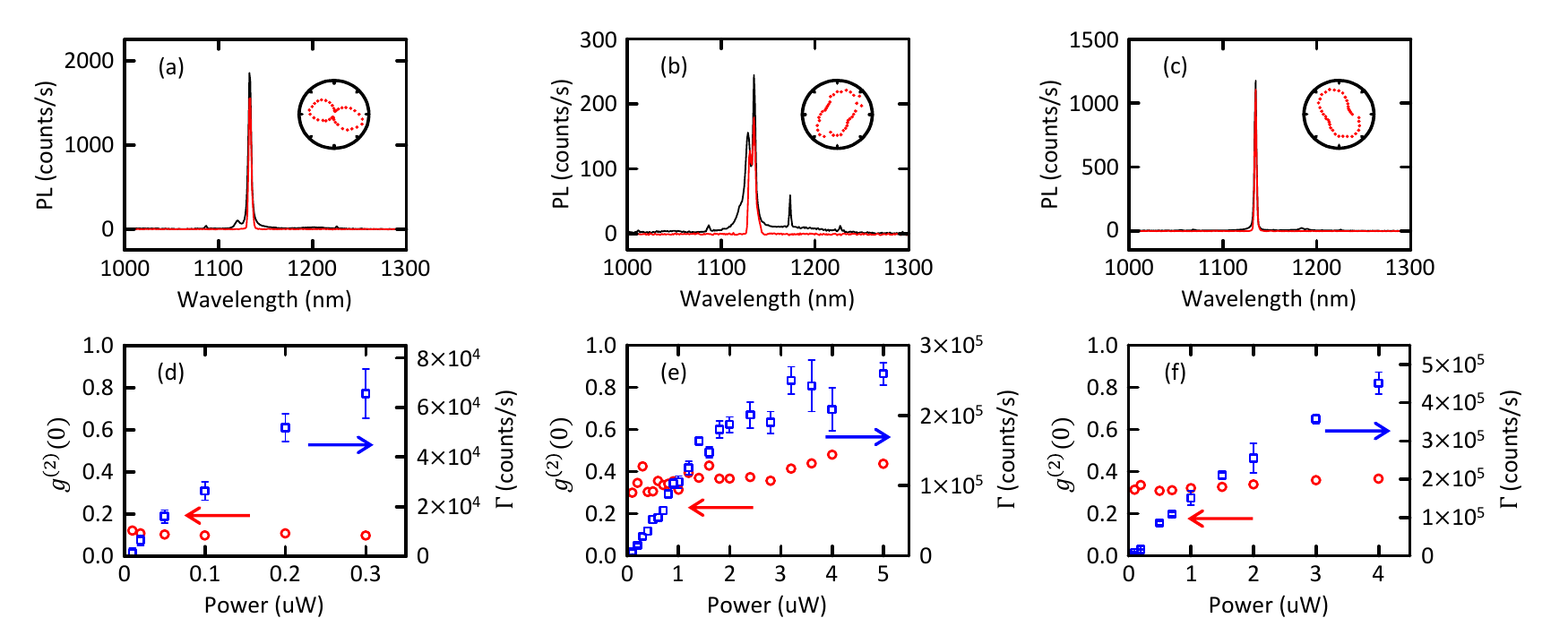}
\caption{
\label{Fig3} (a-c) PL spectra of three different devices taken on the cavities before (black) and after (red) the band-pass filter is set. The filter is tuned to the highest intensity peaks, where we assign the modes at (a) 1133~nm to 2nd mode, (b) 1135~nm to 2nd mode, and (c) 1135~nm to 5th mode. Insets show the laser polarization dependence of the PL intensity for each peak. CW laser with $P=1$~$\mu$W is used for excitation. (d-f) Excitation power dependence of $g^{(2)}(0)$ (red circles) and $\Gamma$ (blue squares) on the cavities measured in (a-c). Pulsed laser is used for excitation. Error bars are the standard deviation of $\Gamma$ obtained by analyzing the time-trace data for each data point. For $g^{(2)}(0)$, error bars are not shown as they are smaller than the symbols in almost all of the data points. For (a) and (d), $X$-polarized laser is used, while $Y$-polarized laser is used for (b, c) and (e, f).}
\end{figure*}

Finally, we investigate excitation power dependence of photon emission statistics on three other devices. The PL spectra on the cavities with and without the band-pass filter are shown in Figs.~\ref{Fig3}(a-c). For these spectrally filtered peaks, we measure $g^{(2)}(0)$ and $\Gamma$ [Figs.~\ref{Fig3}(d-f)] while increasing $P$ until $\Gamma$ shows a rapid drop, which indicates deterioration of the devices. As $P$ increases, $\Gamma$ increases linearly while $g^{(2)}(0)$ remains almost constant, except for the high power region in Fig.~\ref{Fig3}(e), where $\Gamma$ saturates and $g^{(2)}(0)$ slightly increases. In all devices, $g^{(2)}(0)$ remains lower than 0.5 throughout the range of $P$, indicating the robustness of the quantum light signature. This behavior parallels the previous report \cite{He:2017} where defect states for CNTs on a polymer film can also show excellent $g^{(2)}(0)$ values even at relatively high pump powers. At $P=4$~$\mu$W in Fig.~\ref{Fig3}(f), we obtain the highest $\Gamma$ of $4.5 \times 10^5$~counts/s. We note that clear bunching is observed when $P$ is high [Supporting Information S3], which may be caused by an increase of background signals from CNTs which are not coupled to the cavity mode.

In the four devices for which we have measured the photon statistics, we find a positive correlation between single-photon purity and the degree of polarization $\rho$, which is defined by
\begin{equation}
\rho = \frac{I_\mathrm{max}-I_\mathrm{min}}{I_\mathrm{max}+I_\mathrm{min}},
\end{equation}
where $I_\mathrm{max}$ and $I_\mathrm{min}$ are the highest and lowest PL intensity, respectively, obtained by fitting the excitation polarization dependence to a sine function [insets of Fig.~\ref{Fig1}(d) and Figs 3(a-c)]. For the two devices whose $g^{(2)}(0)$ are shown in Fig.~\ref{Fig2}(c) and Fig.~\ref{Fig3}(d), we obtain $g^{(2)}(0) \sim 0.1$ and $\rho \sim 0.8$, while $g^{(2)}(0) \sim 0.35$ and $\rho \sim 0.6$ are obtained for the other two devices. This correlation is reasonable, because low $\rho$ implies that the CNT axis and the localized guided mode polarization do not match or that multiple CNTs with different orientations are coupling to the same mode of the cavity. This observation suggests that controlling the CNT density and orientation on the cavities is a key factor to obtain high quality single photon emitting devices.

The obtained values of $\Gamma$ can be converted to actual photon emission rates at the devices using the total photon collection efficiency in our optical system, which is estimated to be $\sim$2.6\% [Supporting Information S4]. For the highest photon detection rate $\Gamma = 4.5 \times 10^5$~counts/s in our measurements, we obtain the corresponding photon emission rate of $\sim 1.7 \times 10^7$~photons/s. By considering the laser pulse repetition rate of $7.6 \times 10^7$~Hz, the photon emission rate corresponds to the single photon emission efficiency of $\sim$22\%, which is consistent with the estimated quantum efficiency by the lifetime shortening observed in time-resolved measurements. Compared to the previously reported value for similar aryl-functionalized CNTs but wrapped by PFO-bpy and deposited onto a Au-deposited substrate with a separation layer of 160-nm-thick polystyrene \cite{He:2017}, the single photon emission efficiency is almost two times higher.

For achieving further improvement of our devices, optimization of CNT concentration is a key factor as mentioned above. Lowering the CNT density down to an individual CNT level will produce an ideal situation for cavity coupling, but such low density of CNTs results in extremely low yield of cavity-coupled devices. Once appropriate conditions for CNT deposition are determined, spin-coating can be used to obtain more uniform and reproducible deposition of CNTs on cavities \cite{Watahiki:2012}, which enables fabrication of integrated quantum light emitters on silicon chips. As another approach, position-controlled limited-area deposition using a micropipette or nano-droplet \cite{Galliker:2012, Kress:2014} may yield better results because it does not degrade the cavity quality, although such small-volume CNT deposition only at the cavity positions is challenging in practice. Improvement of cavity-coupling efficiency by avoiding quenching from substrates may be possible by using a thinner and more efficient separation layer, like hexagonal boron nitride thin films \cite{Kim:2012}. In addition, the relationship between coupling efficiency and cavity modes is worth investigating, where quality factor and mode profile are different depending on the mode order \cite{Chalcraft:2007, Fujita:2008}. Larger mode volumes are beneficial for obtaining coupling to CNTs on the substrate but results in a lower Purcell effect at the same time. Finally we comment on the tunability of our devices. The emission wavelength of aryl functionalized CNTs can be tuned by selecting chiralities and dopant species \cite{Hartmann:2016, He:2017}, and photonic crystal microcavities have a high flexibility both for absorption and emission resonances \cite{Liu:2015}. Our approach should therefore lead to bright single photon emitters at 1550~nm. In principle, it should also be possible to obtain indistinguishable single photon sources at room temperature by using higher quality cavities.

In summary, we demonstrate integration of carbon nanotube dopant state emitters with silicon microcavities, and PL characteristics and photon statistics of the devices are investigated. PL intensity enhancement by a factor of $\sim$100 is observed from the dopant state emission coupled to the cavity mode, and time-resolved measurements reveal a $\sim$30\% lifetime shortening by the Purcell effect on the cavity-coupled emission. Photon correlation measurements are performed on the devices, and we confirm that room-temperature single photon emission capability, a key feature of $sp^3$-doped CNTs, is preserved in the cavity-enhanced PL emission. We obtain $g^{(2)}(0)$ as low as 0.1 and find that the degree of photon antibunching is stable over a wide range of excitation power. By increasing the excitation power, we obtain a single photon detection rate as high as $4.5 \times 10^5$~Hz, which corresponds to a single photon emission rate of $\sim 1.7 \times 10^7$~Hz and a single photon emission efficiency of $\sim$22\% per laser pulse. Our results indicate that integration of dopant state emitters in CNTs with silicon microcavities can provide bright and high-purity quantum light sources at room temperature on silicon photonics platform, raising expectations toward integrated quantum photonic devices.

\section*{Supporting Information}
See supporting information for autocorrelation histograms taken with and without the band-pass filter, derivation of $\Gamma$, excitation power dependence of autocorrelation histograms, and estimation of photon collection efficiency of the system.

\begin{acknowledgments}
Work supported by JSPS (KAKENHI JP16K13613, JP17H07359) and MEXT (Photon Frontier Network Program, Nanotechnology Platform). A.I. acknowledges support from MERIT, and H.M. is supported by RIKEN Junior Research Associate Program. We thank Advanced Manufacturing Support Team at RIKEN for their assistance in machining. This work was conducted in part at the Center for Integrated Nanotechnologies, a US Department of Energy, Office of Science user facility and supported in part by Los Alamos National Laboratory Directed Research and Development funds.
\end{acknowledgments}

\end{document}